\input harvmac.tex
\Title{\vbox{\baselineskip12pt\hbox{hep-th/9803119}
\hbox{RU-98-06}}}
{\vbox{
\centerline{Anti De Sitter Geometry And}
\vskip 10pt
\centerline{Strongly Coupled Gauge Theories}
}}
\centerline{ Jaume Gomis}
\vskip 15pt
\centerline{\it Department of Physics and Astronomy}
\centerline{\it Rutgers University }
\centerline{\it Piscataway, NJ 08855--0849}
\medskip
\centerline{\tt jaume@physics.rutgers.edu}
\medskip
\bigskip
\noindent
We propose supergravity duals to non-trivial fixed points of the
renormalization group in three dimensions with ADE global symmetries. All the
 fixed point symmetries are identified with space-time symmetries.

\bigskip

\Date{March 1998}
\newsec{Introduction}

Maldacena \nref\juan{J.M. Maldacena, hep-th/9711200.}%
\juan\ has recently conjectured a
very interesting duality relation between 
conformal field theories and space-time 
backgrounds\foot{The near horizon geometry  correspondence
 with branes appeared in  
\nref\sken{K. Sfetsos, K. Skenderis, hep-th/9711138.}%
\nref\frondsal{S. Ferrara, C. Frondsal, hep-th/9712239.}%
\nref\fronds{S. Ferrara, C. Frondsal, hep-th/9802126.}%
\nref\skena{H.J. Boonstra, B. Peeters, K. Skenderis, hep-th/9801076.}%
\nref\CRT{P. Claus, R. Kallosh, A. Van Proeyen, hep-th/9711161,\parskip=0pt
\item{}R. Kallosh, J. Kumar, A. Rajaraman, hep-th/9712073,\parskip=0pt
\item{}P. Claus, R. Kallosh, J. Kumar, P. Townsend, A. Van Proeyen,
hep-th/9802042.}%
\nref\stro{A. Strominger, hep-th/9712251.}%
\nref\hyun{S. Hyun, hep-th/9802026.}%
\nref\juanb{N. Itzhaki, J.M. Maldacena, J. Sonnenschein,
S. Yankielowicz, hep-th/9802042.}%
\nref\minic{M. Gunaydin, D. Minic, hep-th/9802047.}%
\nref\bala{V. Balasubramanian, F. Larsen, hep-th/9802198.}%
\nref\caste{L. Castellani, A. Ceresole, R. D'Auria, P. Fr\'e, S. Ferrara,
M. Trigiante, hep-th/9803039.}%
\refs{\sken,\frondsal,\fronds,\skena,\CRT,\stro,\hyun,\juanb,\minic,\bala,\caste}.}. 
The  precise equivalence between these rather different
theories has not yet been spelled out. An important test of the
duality comes from comparing the symmetries of the field theories with
those of space-time
\nref\witten{E. Witten, hep-th/9802150.}%
\nref\berkooz{M. Berkooz, hep-th/9802195.}%
\refs{\witten,\berkooz}. In 
\nref\Poli{S.S. Gubser, I.R. Klebanov, A.M. Polyakov,
hep-th/9802109.}%
\refs{\Poli,\witten}\ a conjectured equivalence between correlation
functions of the conformal field theory and supergravity actions in
$AdS$ backgrounds is given.
Calculations supporting this correspondence can be found in 
\nref\calc{S.S. Gubser, I.R. Klebanov, A.W. Peet, Phys.Rev. {\bf D54}
(1996) 3915, hep-th/9602135,\parskip=0pt
\item{}S.S. Gubser, I.R. Klebanov, Phys.Lett. {\bf B413} (1997) 41,
hep-th/9708005,\parskip=0pt
\item{}S.S. Gubser, A. Hashimoto, I.R. Klebanov, M. Krasnitz,
hep-th/9803023,\parskip=0pt
\item{}I.Ya. Aref'eva, I.V. Volovich, hep-th/9803028.}%
\calc. In this note we identify the symmetries between 
three dimensional strongly coupled gauge theories and M-theory
backgrounds. These backgrounds have an  $AdS_4\times K$ geometry,
where $K$ is a compact seven dimensional manifold\foot{In section 2 the
geometry will  be made precise.} and correspond to the near horizon
geometry of $N$ M2-branes at an ADE singularity. 
Related theories to those we
analyze have  recently been 
considered in  
\nref\KS{S. Kachru, E. Silverstein,  hep-th/9802183.}%
\nref\Vafa{A. Lawrence, N. Nekrasov, C. Vafa, hep-th/9803015.}%
\nref\Ferr{S. Ferrara, A. Zaffaroni, hep-th/9803060.}
\nref\ber{M. Bershadsky, Z. Kakushadze, C. Vafa, hep-th/9803076.}%
\refs{\KS,\Vafa,\Ferr,\ber}. 

The results of \refs{\fronds,\witten}, are essential to get complete agreement
between symmetries. In \refs{\fronds,\witten} it was found that gauge symmetries in 
the $AdS$ space-time background correspond to global symmetries of the
theory on the boundary of $AdS$. This will enable us to relate the ADE
global symmetry group of the strongly coupled gauge theory with ADE
gauge symmetry in space-time. 

In section two we briefly discuss the three dimensional gauge theories
whose non-trivial fixed points  have
global ADE symmetries. We realize these theories as brane
configurations and identify the space time theory corresponding to the
fixed point to be that of $N$ M2-branes at an ADE singularity. We
carefully analyze the near horizon geometry of such membrane
configurations and find out that it contains an $AdS_4$ part
and that the manifold $K$ can be written as a $S^3$ fibration over an
ADE quotiented  four disk $D_4$. In section three we map  all the
symmetries. Section 4
contains conclusions.

\newsec{Interacting fixed points and space-time geometry}

Three dimensional gauge theories are  asymptotically free and
can be at non-trivial fixed points of the renormalization group in the
infrared\nref\Se{N. Seiberg, Phys.Lett. {\bf B384} (1996) 81,
hep-th/9606017.} \Se. It is believed that non-trivial
fixed points of the 
renormalization group are not only scale invariant but  also conformal
invariant. We will be interested in  three dimensional ${\cal N}=4$
gauge theories (theories with 8 real supercharges) which have
non-trivial fixed points in the infrared with ADE global
symmetries. These conformal field theories exhibit universality since
they have dual short distance descriptions. 
 These theories are related by three dimensional mirror
symmetry\nref\IS{K. Intriligator, N. Seiberg, Phys.Lett. {\bf B387}
(1996) 513, hep-th/9607207} \IS. 

These gauge theories can be realized on branes in string theory. One
of the mirrors can be constructed \foot{When $\Gamma$ acts in the
regular representation.} as the theory on $N$ D2-branes of Type IIA
sitting at a $C^2/\Gamma$ singularity, where $\Gamma$ is an ADE
discrete subgroup  of $SU(2)$
\nref\DM{M.R. Douglas, G. Moore, hep-th/9603167.}%
\nref\JM{C.V. Johnson, R.C. Myers, Phys.Rev. {\bf D55} (1997) 6382,
hep-th/9610140.}%
\refs{\DM,\JM}. The content of the gauge theory can
be conveniently summarized by a quiver diagram.  The gauge group can be read
from the nodes of the extended Dynkin diagram of the corresponding ADE
Lie algebra. The 
hypermultiplets are given by the links of the extended Dynkin
diagram and are in the bifundamental representation of the adjacent
nodes\foot{See \refs{\DM,\JM}\ for more details.}. These theories have
a quantum moduli space of vacua with two 
branches. In \Se\ these theories were solved using string theory
considerations and in 
\nref\BHO{J. de Boer, K. Hori, H. Ooguri, Y. Oz, Z. Yin,
Nucl.Phys. {\bf B493} (1997) 148, hep-th/9612131.}%
\refs{\IS,\BHO} they were analyzed by field theory methods. The
Coulomb branch is the  
moduli space of instantons of the corresponding ADE gauge group. The Higgs
branch, obtained as a  hyper-Kahler quotient, is an ADE singularity. The two
branches intersect at the origin of the ADE singularity, and the theory there
 is at a non-trivial fixed point of the renormalization
group with ADE global symmetry \Se. It is important  to point
out that the nontrivial fixed point is in the strongly coupled regime
$g^2_{YM}\rightarrow \infty$ \Se. 

The space-time theory corresponding to this fixed point is
modified. To tune to the fixed point one needs 
to send the Yang-Mills coupling constant to infinity. By open string
considerations  $g^2_{YM}\propto g_s$, where $g_s$ is the coupling
constant of the Type IIA theory on which the D2-branes are
embedded. Thus, to tune to the fixed point we must push the space-time
picture to M-theory. Therefore, the space-time description of the
fixed point is that of N M2-branes sitting at an ADE singularity. 

Following the AdS-CFT correspondence we wish to identify all the
symmetries of the three dimensional interacting fixed points with
those of M-theory on  the near horizon geometry of $N$ M2-branes on an
ADE singularity. In order to identify the near horizon geometry
corresponding to this brane configuration it is instructive to write
down the near horizon geometry of M2-branes in flat space. The 
 metric is given by
\eqn\solu{
ds^2={r^4\over R^4}d\vec{x}^2+{R^2\over
r^2}dr^2+R^2d\Omega_{7}^2,} where $R^6=2^5\pi^2Nl_p^6$ and
$r^2=x_3^2+\ldots x_{10}^2$\foot{The membranes lie in the 012
plane, so that  the coordinates $x_i, i=3,\ldots 10$ 
are transverse to the membranes.}.  The geometry is that of $AdS_4\times
S^7$. The original conjecture by Maldacena is that the three
dimensional conformal field theory corresponding to the theory of $N$
M2-branes is dual to M-theory on $AdS_4\times S^7$. The conformal group
$SO(3,2)$ 
is identified with the Anti de Sitter group in space-time and the
$SO(8)$ R-symmetry group of the field theory with the isometry group
of $S^7$.

We want to find  the near horizon geometry of N M2-branes at
a $C^2/\Gamma$ singularity. For concreteness we will concentrate on
the case $\Gamma=A_{k-1}=Z_k$, but the generalization to the other ADE
discrete groups is straightforward.
 Let $z_1,z_2$ be the complex coordinates of the $C^2/Z_k$
singularity. The orbifold action is given by
\eqn\orb{\eqalign{
z_1&\rightarrow e^{2\pi i\over k}z_1\cr
z_2&\rightarrow e^{-{2\pi i\over k}}z_2.}}
This action leaves the $AdS_4$ geometry invariant but it acts
non-trivially on the $S^7$ geometry. Let us analyze carefully the
orbifold action on $S^7$. Consider\foot{The following geometrical
analysis was done in collaboration with D.-E. Diaconescu.}
$S^7\subset C^4$
defined by  
\eqn\sphere{
|z_1|^2+|z_2|^2+|z_3|^2+|z_4|^2=a^2.}
The $Z_k$ orbifold group acts 
as in \orb\ in the $z_1,z_2$ coordinates and leaves 
 $z_3,z_4$ invariant. Let $\pi:C^4\rightarrow C^2(z_1,z_2)$ 
denote the
projection onto the $(z_1,z_2)$ plane. Then the restriction
${\pi}|_{S^7}$ exhibits $S^7$ as an $S^3$-fibration over the disk
$D_4$ of radius $a$ in $C^2(z_1,z_2)$.
The radius of the $S^3$ fiber varies over the base such that 
it shrinks to zero size at the boundary $\del D_4$. The fibration is 
not globally trivial. The orbifold action identifies the $S^3$ 
fibers over a $Z_k$ orbit through any arbitrary point $(z_1,z_2)\neq
(0,0)$. At the same time, it leaves the $S^3$ fiber over the origin 
invariant. Therefore the quotient is an $S^3$ fibration over the
singular space $D_4/Z_k$. Note that the singular locus of the total
space consists of an $A_{k-1}$ singularity fibered over the central 
$S^3$ fiber. 

 Now that the near horizon geometry has been
found we can identify the symmetries of three dimensional fixed
points with those of M-theory on $AdS_4\times S^3\times_{f} D_4/\Gamma$. 

\newsec{Space-Time Symmetries and Fixed Point Symmetries}

The identification of the conformal field theory symmetries with the space-time
symmetries can be done as follows. Since the field theory is at a 
non-trivial fixed point of the renormalization group the scale
symmetry is enlarged to the conformal group in three dimensions,
$SO(3,2)$. This can be identified with the Anti de Sitter symmetry
group in the space-time theory. For this it is
 crucial for the
orbifold action not to affect the $AdS_4$ geometry. The gauge theories
we started with 
have  ${\cal N}=4$ supersymmetry in three dimensions, and the
$SO(4)$ R-symmetry group can be
identified with the isometry group of the  $S^3$ fiber. 

Moreover, the CFT has an ADE global symmetry arising at  the origin of
the moduli space of ADE instantons. As shown in \refs{\fronds,\witten} global
symmetries in the CFT correspond to gauge symmetries in the
$AdS$ geometry. We will show that the ADE global symmetry can be
identified with ADE gauge symmetry of M-theory on $AdS_4\times
S^3\times_f D_4/\Gamma$. In M-theory we can wrap membranes on the
shrunken cycles of the ADE singularity. This gives rise to  ADE 
gauge bosons in the remaining directions. In order to be left with
ADE gauge symmetry in $AdS_4$ we need to reduce the gauge fields on
$S^3$, and since dim$(H^0(S^3))=1$, one is left with ADE gauge
bosons in $AdS_4$. This ADE gauge symmetry is the one we want to
identify with the ADE global symmetry of the fixed point. Therefore, by
analyzing the space-time geometry corresponding the fixed point we have
identified all the symmetries in the two theories.

\newsec{Conclusions}

A preliminary step towards understanding the AdS-CFT correspondence is
to identify the symmetries of both theories. In this note we do that
for three dimensional conformal field theories with ADE global
symmetries by identifying the
space-time description of these fixed points. Some of the symmetries
of the field theory arise as geometrical symmetries of the near
horizon geometry. The ADE global symmetries correspond to ADE gauge
symmetries in space-time.

It would be interesting to perform similar test for other non-trivial
fixed points of the renormalization group. Eventually, one might hope
to be able to describe all such fixed points by theories in
space-time. Along the lines of 
\nref\conformal{O. Aharony, Y. Oz, Z. Yin, hep-th/9803051,\parskip=0pt
\item{}S. Minwalla, hep-th/9803053,\parskip=0pt
\item{}R.G. Leigh, M. Rozali, hep-th/9803068,\parskip=0pt
\item{}E. Halyo, hep-th/9803077.}%
\refs{\Poli,\witten,\conformal} it would be
interesting to compute some correlation functions  and anomalous
dimensions of these conformal
field theories from M-theory considerations. 

\vfill\eject
After this work was completed the paper  
\nref\tert{S. Ferrara, A. Kehagias, H. Partouche, A. Zaffaroni,
hep-th/980310.}%
\refs{\tert} appeared with similar subject.
\centerline{\bf Acknowledgments}
I would like to thank D.-E. Diaconescu, M.R. Douglas, B. Fiol and
S. Shenker for discussions.  

\listrefs

\end